# Retinal and post-retinal contributions to the Quantum efficiency of the human eye.

Authors: Gibran Manasseh[1*], Chloe de Balthasar[1], Bruno Sanguinetti[2], Enrico Pomarico[2], Nicolas Gisin[2], Rolando Grave de Peralta[1], Sara L. Gonzalez[1]

[1]Electrical Neuroimaging Group, Department of Clinical Neuroscience, Faculty of Medicine, Switzerland

[2]Group of Applied Physics, Faculty of Physics, University of Geneva, Switzerland

Correspondence :
Gibran Manasseh
University of Geneva
Department of Clinical Neuroscience
Faculty of Medicine
Rue Gabrielle-Perret-Gentil, 4
1211 Geneva 14
Switzerland

Abstract: The retina is one of the best known quantum detectors with rods able to respond to a single photon. However, estimates on the number of photons eliciting conscious perception, based on signal detection theory, are systematically above these values. One possibility is that post-retinal processing significantly contributes to the decrease in the quantum efficiency determined by signal detection. We carried out experiments in humans using controlled sources of light while recording EEG and reaction times. Half of the participants behaved as noisy detectors reporting perception in trials where no light was sent. DN subjects were significantly faster to take decisions. Reaction times significantly increased with the decrease in the number of photons. This trend was reflected in the latency and onset of the EEG responses over frontal and parietal contacts where the first significant differences in latency comparable to differences in reaction time appeared. Delays in latency of neural responses across intensities were observed later over visual areas suggesting that they are due to the time required to reach the decision threshold in decision areas rather than to longer integration times at sensory areas. Our results suggest that post-retinal processing significantly contribute to increase detection noise and thresholds, decreasing the efficiency of the retina brain detector system.

## Introduction

The first experiments on the sensibility of the human eye to weak, near absolute thresholds, optical signals were conducted in the 1940s (Hecht, 1942). They led to the conclusion that rod photoreceptors can detect a very small number of photons, typically less than 10 during an integration time of about 300 ms (Barlow, 1956). This prediction has been confirmed by several experiments (Rieke and Baylor, 1998) making from the human eye a remarkable light sensitive detector, which can easily stand a comparison to today's best man-made detectors (Rieke and Baylor, 1998). This has even led to the proposal of using the human eye as a detector for quantum phenomena such as entanglement (Brunner, 2008;Sekatski et al., 2009).

The quantum efficiency (QE) of the human eye as a detector, i.e., the probability of getting a response given that a photon impinges on the retina, has been determined into two different ways: behaviorally and by direct neural recordings. In behavioral terms the QE can be estimated from the frequency of seeing curves (Foes) (Hecht, 1942) later replaced by a distribution of ratings (Sakitt, 1972) . Flashes of light, with a controlled probabilistic distribution of photons are sent into the pupil and subjects who are dark adapted are prompted to indicate if they perceived a flash. The detection threshold, i.e., the number of photons required to trigger a conscious percept (arbitrarily defined as the light intensity giving rise to 60% detection), is determined by measuring the fraction of trials in which a flash is reported as perceived as a function of the number of photons incident at the cornea. Since, only about 8% of the photons incident on the cornea reach the retina, hence about 100 photons are required to trigger a neural response even if rod photoreceptors can react to single photons (Rieke and Baylor, 1998).

Direct neural recordings have been used in toads (Baylor, 1979) and monkeys (Baylor, 1984) to determine QE from the number of photons needed to evoke responses in isolated rod photoreceptors. These studies lead to the conclusion that rod photoreceptors can signal the absorption of single photons. Consequently, estimates of QE vary in about one order of magnitude as a function of the definition of response (neural response in photoreceptors vs. behavioural responses) used for its quantification. Behavioural measurements based on the FoS curve place the quantum efficiency (QE) of the human eye between 0.03 and 0.06 while direct estimates based on losses within the eye range from 0.1 to 0.3 (Baylor, 1979). Consequently, the QE estimated from behaviour is very low compared with the absorptive QE estimated from the properties of light photoreceptors at the retina.

The reasons for the divergence in the estimated and measured QE are not completely clear as the processes limiting sensitivity are not yet fully characterized. When estimating QE from the FoS curve we demand to the observer to indicate whether or not they perceived the stimuli. According to classical psychophysical models (Krantz, 1969), this detection process is composed of at least two psychological components or processes: 1) the sensory process transforming the physical stimulation into internal sensations and 2) a decision process which decides on responses based on the output of the sensory process. Each of the two processes (see Figure 1) is, in turn, characterized by at least one parameter: The sensory process by a sensitivity parameter and the decision process by a response criterion parameter. To avoid confounding the sensitivity of the sensory process with the response criterion of the decision process, one needs to measure two aspects of detection performance: the conditional probability that the observer says “yes” when a stimulus is present (the hit rate, or True positive rate closely linked to the FoS curve) but also the conditional probability that the observer says “yes” when the stimulus is not present (False positive rate or FAR).

Barlow (Barlow, 1956;Hallett, 1969) relied on the concept of false positive responses to explain the discrepancy in QE. He attributed the fact that observers occasionally report seeing a flash even when no light was delivered to the existence of what he termed “dark light” or “dark noise”. Behavioural sensitivity and dark noise can be the result of Poisson fluctuations in photon absorption at the level of the retina. Experiments in toads have shown that one of the possible source of this dark-noise is the thermal ionization of the photosensitive protein in the retina, pointing that sensitivity of frog was decreasing with temperature (Aho et al., 1993). Upon this model, dark noise increases the rate of false-positive affecting the sensitivity of the detection threshold and the reliability of the QE estimated from behavior.

Is retinal noise the only factor impacting the response criterion that characterizes the decision process? If this were the case then retinal (dark) noise would be the only explanation for the observed discrepancy in QE. Yet, noise is not an exclusive property of retinal photoreceptors. Noise might arise anywhere into the chain of neural processing and add to fluctuations at the level of the retina. Supporting the existence of postretinal contributions to dark noise are the multiple experiments that rely on Transcranial Magnetic Stimulation (references) at the level of the occipital cortex to evoke phosphenes. Some of these studies indicate that the phosphene perception appears after extensive recurrent processing and is therefore not purely attributable to primary visual processing (Taylor et al., 2010). In addition, deciding that the stimulus is present or absent is clearly not a matter of sensory evidence alone as a decision about stimulus absence lacks sensory evidence by definition. Interestingly, cells coding for the decision about both, the absence and presence of stimulus, have been recently reported at the prefrontal cortex of primates (Merten and Nieder, 2012). Noise at the level of prefrontal cells might equally impact decisions contributing to the discrepancy in QE. The impact of post-retinal processing on the reduction of QE remains unknown.

To shed further light on the post-retinal mechanisms impacting the quantum efficiency of the human eye we repeated a version of the FoS experiment described in (Hecht, 1942). We hypothesized that post-retinal processes substantially contribute to the observed decrease in QE when inferred from the FoS curve. We adhered to the only accessible measure of QE we can rigorously control in this experiment, i.e., the fraction of incident photons that contribute to conscious perception as measured by the FoS curve. Besides psychophysics and signal detection theory, we relied on two other complementary methodologies helpful to dissociate the stages of processing impacting sensitivity (see Figure 1). First, reaction times (mental chronometry) which allows inferring to some extent the content, duration, and temporal sequencing of cognitive operations within perceptual processes (Jensen, 2006). Second, scalp measured Event related Potentials (ERPs) which provide an indicator of the latency of neural responses at the different processing stages (Thorpe et al., 1996).

## Material and Methods

### 1. Recording Protocol

*Participants*

Twelve healthy young volunteers (age range: 26-38, mean age 30±4, 2 females) were recruited from the faculties of Physics and Medicine of the University of Geneva. Eleven of the participants were right-handed. They had no history of neurological problems. The whole experiment was approved by the local ethics committee (Geneva University Hospitals). Participants were verbally informed of the goals of the experiment and the sequence of events.

*Dark adaptation*

The experiment was carried out in a completely dark recording room with all potential sources of light, e.g., computer LEDS, covered by black plastic tape. Participants were dark-adapted before the experiment by being kept during 40 minutes in the dark room while wearing a black sleeping mask. The darkness was kept during the approximately two hour's duration of the experiment.

*Rationale of the experiment and instructions to participants*

As illustrated in Figure 1, we don't have direct access to the sensory and decision processes we would like to disentangle in this study. We therefore need to infer the contribution of each process trough the analysis of observable measures of neural activity such as the EEG or from behavior (using for example Signal detection theory and RTs). We expected increases in sensitivity but decreases in RTs with stimulus intensity. While sensitivity is generally a relatively stable property of the sensory process across subjects, the decision criterion can vary widely from subject to subject and from time to time. We then expected to capture information about the interindividual variability and sources of noise linked to decisions trough the investigation of reaction times. By combining information from reaction times and the latency and scalp positions of peaks in the ERP signal we expected to extract information on the relative timing of the sensory and decision processes. According to current decision making models, decisions are taken when sufficient evidence in favor of one of another choice has been accumulated. Decisions occur when neural signals reach a certain threshold. Noise alters the gradual accumulation of neural information speeding up or retarding the decision time. Consequently, one should expect that differences in peak latencies which correlate with differences in RTs across stimulus intensities appear above the areas where neural responses are being accumulated. Depending on whether the differences appear over sensory or decision areas one can then infer the timing of their relative contributions.

The link between the FoS and RTs can be expressed as:

$$FoS = f(I) \leftrightarrow I = f^{-1}(FoS)$$
$$RT = g(I) \leftrightarrow RT = g\left(f^{-1}(FoS)\right) = h(FoS)$$

where I is the number of photons sent. According to this expression, the RT is 1) a function of the frequency of seeing (FoS) and 2) dependent on the decision process.

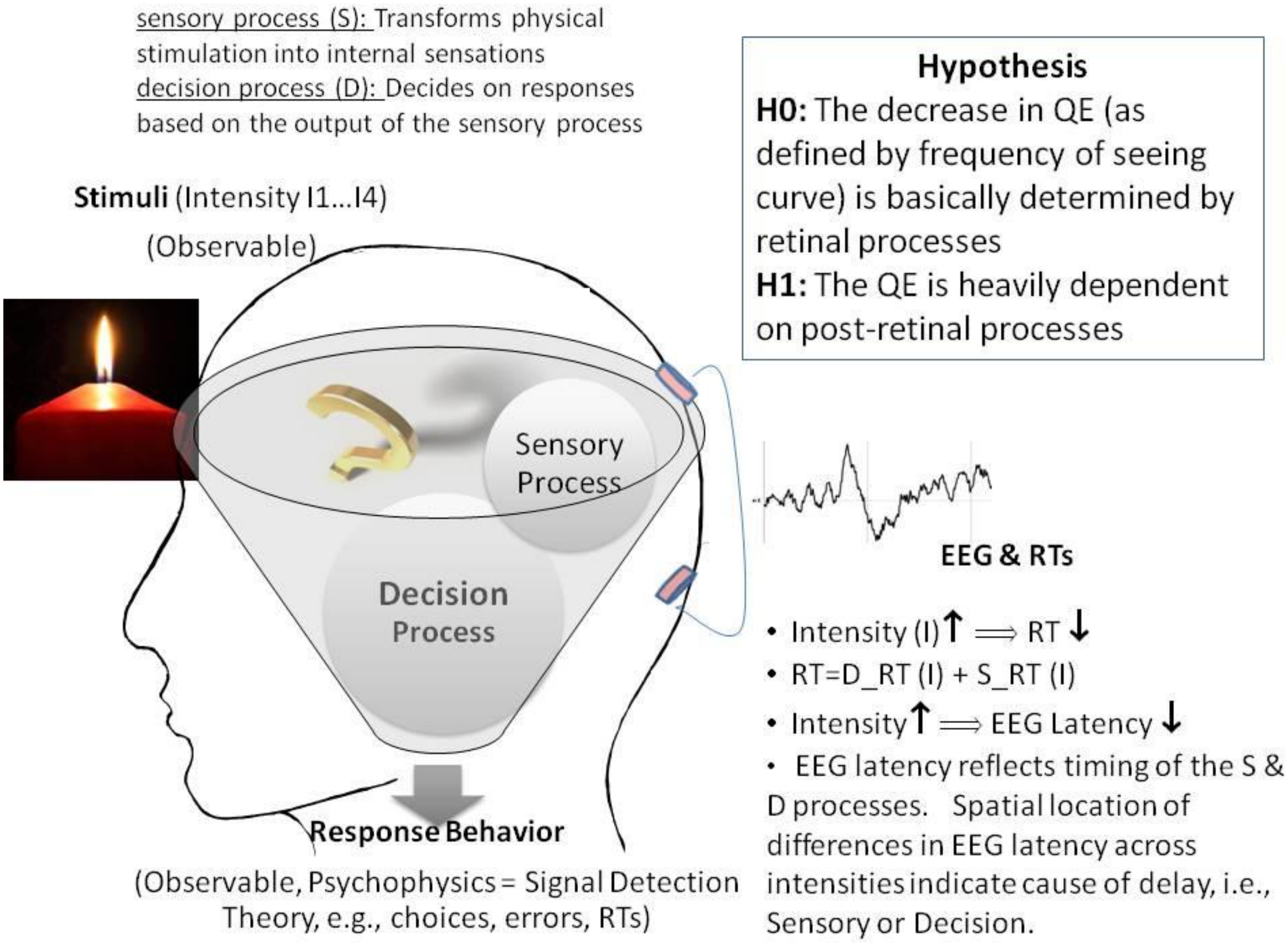

**Figure 1**

The instruction given to the subjects was explicit: "report seeing the flash via a button press when you feel completely confident about the percept". We avoided a multiple choices task reflecting the trial by trial confidence in perception (Barlow, 1956) & Sakitt (Sakitt, 1972). We did so on three bases: 1) Reaction Times are known to vary as a function of the difficulty of the task. Yet, the difficulty of the task is not only linked to the perceptual difficulty that we want to investigate here but also to the number of choices available. There is ample experimental evidence supporting the increase in reaction times with the number of available choices. Since we were interested in the link between perceptual processes and reaction times rather than on the link between choices and reaction times we considered the two choices alternative as the most reasonable one. 2) The number of errors is known to increase with number of choices as repeatedly shown in the literature. This is sometimes due to false button presses. Since choosing between several buttons in full darkness is more challenging than under normal illumination conditions then the probability of false button presses increases. Consequently, the two choices task adopted here is "optimal" to: (i) isolate real "dark noise” coming from retinal/post-retinal effects from motor mistakes and (ii) to isolate the perceptual component of the reaction times from the choice component. 3) Subject’s performance and EEG signals tend both to worsen with the duration of the experiments. As previously explained, multiple choices tasks lengthen the experiment. A condition for the experiment is to remain attentive and still as to obtain adequate signal to noise ratios in EEG signals and sustained performance. This posed a challenge to some of the participants as the full experiment lasted for approximately 2.30 hours.

*EEG Recordings*

The scalp electroencephalogram (EEG, 64 channels) and reaction times (RTs) were recorded during the experiment. The EEG was recorded at variable frequency sampling (1024 Hz or 2048Hz) to guarantee the temporal precision of the triggers and response time. Frequency sampling was individually selected on the basis of the initial psychometric curves. Recordings were done using the Biosemi system with 64 sintered Ag-AgCl electrodes and implicit filter settings at 5th order sinc filter with a -3 dB point at 1/5th of the sampling frequency. The electrodes were mounted on the manufacturer-provided cap according to an extended 10-20 system. The Biosemi system uses a common mode sense (CMS) active electrode as the reference. Visual inspection was used to reject artifact-contaminated trials. Bad electrodes interpolation was based on spherical splines using Cartool. Epochs of 1100 ms (100 ms before the presentation of the stimulus) were extracted after notch filtering at 50 Hz and superior harmonics. Baseline correction was based on 100ms prestimulus window.

*Experimental Setup*

The experimental design is schematically depicted in Figure 2. A light emitting diode (LED) was used to produce flashes of light at 500nm wavelength which guarantees maximum sensitivity of rod cells (Alpern, 1987). A portion of the light was collimated and coupled into a single mode fiber. This kind of optical source was chosen for safety reasons as only a maximum power of hundreds of pW can be coupled into the fiber. The LED was software controlled via a National Instruments digital to analog card that allows varying the power of each light pulse between 8 pW and 400 pW, while its duration can be chosen between 100µs and 1ms. In this way, the number of photons in each pulse can be dynamically varied by nearly three orders of magnitude. We used neutral density filters (grey filters) to further decrease the optical intensities by a factor $\mathbf{t_{ND}}$ and adapt them to the subject’s sensitivity that was individually detected as described in next section. While $\mathbf{t_{ND}}$ changes from subject to subject it is set to at least 0.1.

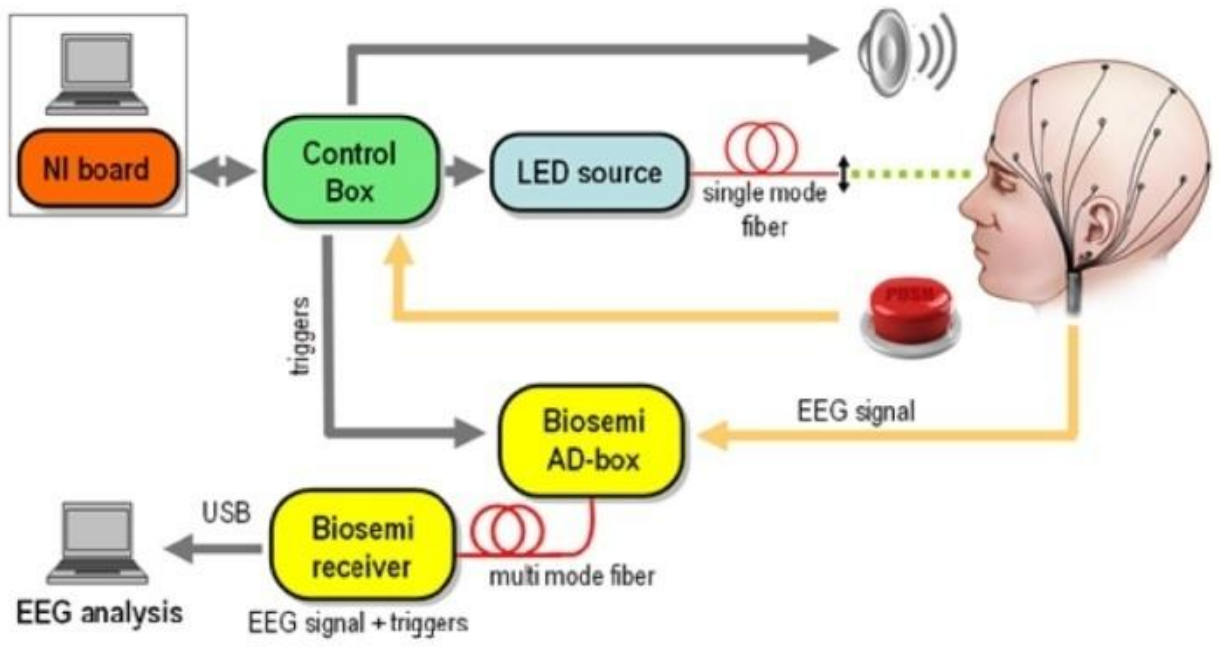

**Figure 2**

The light coupled into the fiber was directed to one eye of the subject, who rested his chin and forehead on a chinrest support to keep the head steady over the experiment. A collimation lens at the end of the fiber allows focusing of the beam on the retina. Since the density of rod cells (the most sensitive human photo-receptors) is highest in the peripheral region of the retina, the direction of the beam is chosen to form an angle of approximately 20 degrees with respect to the eye's axis. The retina is illuminated on the temporal side, to avoid the optical nerve. At least 700 ms before the light is emitted, an acoustic signal is produced to alert the subjects of the imminent emission of the pulses. The subjects press a button in case of conscious perception of the flash and a digital signal is sent to the NI board. The communication with the board is managed via a control box. Finally, the control box sends to the Biosemi AD-Box the trigger signals corresponding to 1) the timing of the acoustic signal, 2) the value of the randomly chosen intensity and 3) the timing of the button press for perceived flashes. At least 150 repetitions of each intensity and the same amount of blank trials (the acoustic signal is given but the flash is not sent) were obtained per each subject. The time course of the whole experiment is shown in Figure 3.

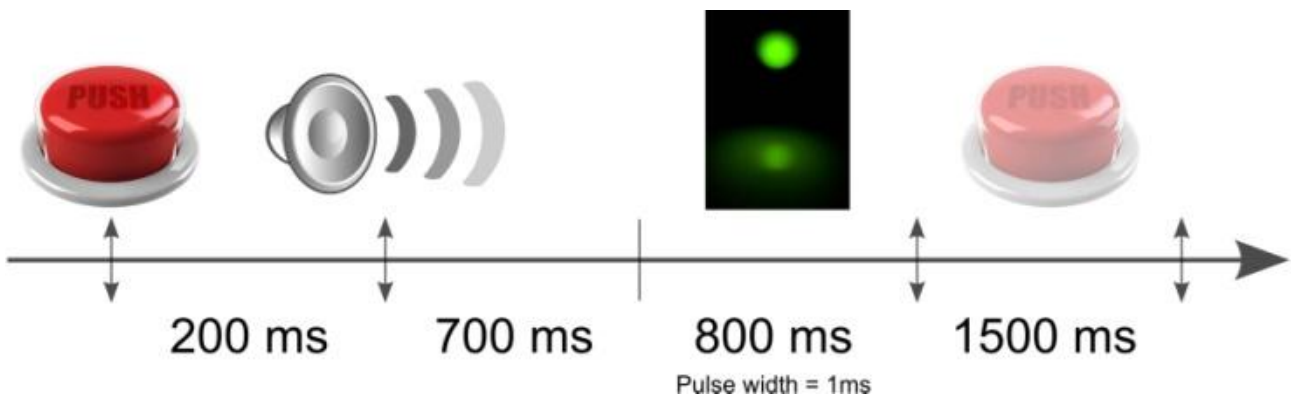

**Figure 3**

*Estimating the Number of photons incident at the cornea as a function of the power emitted by the LED.*

The power emitted by a LED is linear with respect to the voltage bias applied to it and linearity was assessed in our case to hold up to 9 Volts. A calibrated power meter was used to measure the power $\boldsymbol{P}$ which exits the fiber. The number of photons per second corresponding to the power measured when the LED is at 9V is given by:

$$\boldsymbol{n}^* = \frac{\boldsymbol{P\lambda}}{\boldsymbol{hc}} \qquad \textbf{(1)}$$

where $\lambda$ is the wavelength, $\boldsymbol{h}$ is Planck constant and $\boldsymbol{c}$ is the speed of the light. For each subject 4 different intensities were individually selected by varying the voltage applied to the LED from the $V_{offset} \approx 2.5\, V\ to\ 9\, V$. As the power emitted by the LED is linear between 2.5V and 9V, the value of $\boldsymbol{P}$ in equation (1) must be multiplied by a factor $t_{bias} = (V_I - V_{offset})/(9V - V_{offset})$, where $V_I$the voltage is applied to the LED for the specific intensity I.

The fact that the grey filters reduce the power by a factor $t_{ND}$ must be taken into account into the calculations as well as the fact that the pulses had a very short duration $\Delta t_{pulse}$ that varied according to the subject from 0.5 to 1 ms. Therefore, the final formula used to compute the number of photons at each pulse that were sent to the cornea is finally given by:

$$n_{pulse} = \frac{t_{bias} \times t_{ND} \times \Delta t_{pulse} \times P \times \lambda}{h \times c} \quad \textbf{(2)}$$

*Selection of the best individual parameters (Method of Adjustment)*

Once subjects were adapted to darkness and the EEG set up installed we carried out some initial tests to individually tune the attenuation of light (achieved by the grey filters) and the duration of the pulses. Several attenuations were tested for each of the four intensities and subjects were requested to report if they perceived the flashes. The final attenuations were chosen as those for which the flashes were perceived in half of the presented pulses. This approach was preferred to the on-line selection of the intensities in order to minimize the duration of the experiment. Finally, during the experiment, four intensities of flashes were presented to each subject with attenuations chosen to evoke perception in half of the trials.  From now one, we will use the term intensity for the intensity of the beam reaching the cornea, i.e., after the filters.

## 2. Data analysis and rationale

### *2.1 Reaction Time Analysis*

We investigated if and how reaction times, defined as the time elapsed between the onset of the flash and the button press indicating perception, vary as a function of 1) the number of incident photons, and 2) the accuracy of decisions. There is indeed growing evidence from primate neurophysiology indicating that in reaction time tasks, a perceptual choice is made when the firing rate of a selective cortical neural population reaches a threshold (Lo and Wang, 2006). Reaction times (RTs) therefore correlate with the time needed to reach the threshold that is dependent on the difficulty of the choice (Smith, 2004). Weak or uncertain stimuli lead to slowly varying accumulation of evidence and longer decision times while certain/strong stimuli lead to quickly growing accumulation of certitude that is reflected in a sharp build up of neural activity that quickly reaches the necessary threshold to reach decisions speeding up the reaction times. On this basis we should expect: 1) an inverse dependency between reaction times and the intensity (number of photons) of the flashes, 2) Significant differences between mean RTs corresponding to different stimulus intensities. On the other hand, subjects performing a visual detection task need to manage the tradeoff between speed and accuracy (Ratcliff, 2002).

### *2.2 Statistical analysis of Behavioral data*

RTs were compared using the one-way ANOVA if data had a normal distribution (according to the Lilliefors test, matlab statistical toolbox, 0.05 significance level) or using the Kruskal-Wallis test (nonparametric one-way ANOVA) for non-normally distributed data. Unless otherwise specified, all the statistics and analysis were done using Matlab R2006a. When appropriated, we display behavioral data using notched boxplots that provide a summary of several important features of the distribution of values (e.g., median, confidence interval around it, outliers). In these plots, when the notches of two or more groups do not overlap then the medians of the two groups differ at the 5% significance level.

### *2.3 Frequency of Seeing (Accuracy) and QE*

The 4 pre-selected intensities and the blank trials were presented to the subjects in random order with each intensity repeated at least 150 times. The frequency of seeing curve was obtained plotting the proportion of flashes reported as perceived as a function of the intensity. The energy of the flashes was transformed into the average number of photons using equation (2).

The probability of seeing decrease with the decrease in the number of presented photons. However, determining which threshold should be used to determine the minimal number of photons necessary to generate conscious detection is still an open question. Hecht & al set the threshold at 60% of probability and concluded that 54 to 148 incident photons are needed to trigger conscious detection. Here, we decided to set a slightly lower threshold at 50%. This is due to three reasons, 1) we introduced under the form of zero intensity trials a control against dark noise, i.e., detection by chance and 2) we used several naïve (untrained) subjects and 3) the experimental design introduced a variable delay between the acoustic signal and the photons arrival to prevent anticipation.

In order to more precisely estimate the number of flashes absorbed by the retina that are necessary for conscious perception from discrete observations, the probability of seen curve is typically fitted with a model. For instance, Rieke (Rieke and Baylor, 1998), following Hecht (Hecht, 1942), made the assumption that for a given intensity the number of photons absorbed by the retina follows a Poisson-distribution. In Rieke's model, the probability of seeing a flash ($p_{see}$) of intensity $I$ can be written as:

$$\boldsymbol{p_{see}(I) = \sum_{n=\theta}^{\infty} \frac{e^{(-\alpha I)}}{n!}(\alpha I)^n} \qquad \textbf{(3)}$$

Where $\theta$ is the minimal number of photons, under which the subject never perceived a flash and $\alpha$ represents the decrease in intensity between the number of photon sent in the flash and the amount of photons arriving at the retina. The value $\alpha \times I$ represents thus the mean number of photon absorbed by the retina.

However, after extensive testing on the data we found that much better fits are obtained assuming a log-poisson regression model distribution, i.e.,

$$\boldsymbol{p_{see}(I) = \sum_{n=\theta}^{\infty} \frac{e^{(-log(\alpha I))}}{n!} log(\alpha I)^n} \qquad \textbf{(4)}$$

The free parameters $\theta$ and $\alpha$ were therefore determined by the simultaneous minimization of the objective function $f$ given by:

$$\boldsymbol{f(\theta, \alpha) = \sum_{k=1}^{4} \left[ p_{see}(I_k) - \sum_{n=\theta}^{\infty} \frac{e^{(-\log(\alpha I_k))}}{n!} \log(\alpha I_k)^n \right]^2} \qquad \textbf{(5)}$$

Where $I_k$ are the four intensities presented to each subject. Note that the fits are here used to get an adequate approximation of the number of photons needed for conscious detection at the 50% probability. By no way, should these fits be considered as a model for the detection probability as they have been obtained from just four intensities.

### *2.5 Measuring sensitivity using signal detection theory: the sensitivity index (d') and the Receiver Operating Curve (ROC)*

A criticism to the FoS curve as a measure of sensory sensitivity is that it mixes the decision criteria with sensory sensitivity. We therefore used as an additional measure the sensitivity index (d') which splits detection performance into two components: the conditional probability that the observer says "yes" when a light is present (the hit rate or true positive rate TPR) and the conditional probability that the observer says "yes" when a light is not present (the false alarm rate, or false positive rate FN) (Green and Swets, 1966). We computed the sensitivity index d' as:

$$d' = z(FPR) - z(TPR)$$

where $z(\cdot)$ denotes the z-score of the given probabilities, FPR is the false-positive rate and TPR is the true-positive rate. Since $z(0) = \infty$ , we replaced FPR by 1/N, with N denoting the number of trials, for subjects showing perfect specificity.

A different way to test the overall efficiency of the human visual system is to use the ROC curve, a measure coming from signal detection theory (Fawcett, 2006). The ROC curve is the plot of the fraction of true positives (i.e. true positive rate $TPR$) vs. the fraction of false positives (i.e. false positive rate $FPR$) of a binary classifier as its discrimination threshold is varied. A true positive ($TP$) is a non-zero intensity trial reported as perceived while a true negative ($TN$) is a non perceived zero intensity trial. A false negative ($FN$) is a non zero intensity trial that is non perceived and a false positive is a zero intensity trial reported as perceived. The $TPR$ and $FPR$ are respectively defined as:

$$\boldsymbol{TPR = TP/(TP + FN)} \quad \textbf{(6)}$$

$$\boldsymbol{FPR = FP/(FP + TN)} \quad \textbf{(7)}$$

The advantage of the ROC over the FoS is that it allows quantifying both, the sensitivity of the observer given by the TPR as well as his/her specificity defined as 1-FPR.

### *2.7 EEG analysis*

ERPs were computed by averaging the epochs reported as perceived by button presses, once aligned by the flash onset for every subject and intensity. Both, the original Biosemi reference and the average reference were used in the analysis to assess independence of the effects on the chosen reference. Note that the average reference removes the effect of a constant from the data. The Grand Mean (GM) was afterwards computed as the average over subjects of the individual ERPs once normalized by the norm of the global scalp energy. This normalization avoids overweighting the contribution of individual subjects to the GM.

The excellent (millisecond basis) resolution of the EEG makes it suitable to investigate the timing and scalp topography of the neural responses as a function of the variations in the number of perceived photons. The timing of the responses might help to elucidate post-retinal contributions to the QE of the visual system. Indeed, it has been observed that neural activity needs to reach a certain decisional threshold before the stimuli is perceived. Weak stimuli lead to builds up in neural activity characterized by flatter slopes than strong stimuli. Therefore, we should expect that the timing of the component of the ERPs varies as a function of the stimulus intensities over areas involved in perceptual or decisional processes In fact, if perceptual decisions are responsible for the QE of the visual system, delays in the ERPs as a function of the intensity should already appear at the level of the primary visual cortex reflecting the transmission delays at the retinal level. On the other hand, if there are post-retinal contributions to the QE, we should expect additional delays over frontal or parietal areas known to be involved in decisional processes. Consequently, the analysis of ERPs might help us to elucidate the contribution of the different brain regions to the dependency of the reaction times with intensity. Indeed, if delays already occur in the transmission of signals from the retina, delays (differences in the latency of the ERP components) in the ERPs at the level of the primary visual cortex should be observed. On the other hand, if decisional processes are the only responsible for the delays in the behavioral responses, the first delay observed should appear on ERPs of brain areas involved in more complex and cognitive processes, as parietal and prefrontal areas

To evaluate the delays in the mean ERPs responses (as well as for the Grand Mean ERPs) as a function of the intensity we computed for each participant and for each electrode, the pair-wise time-lagged Spearman's rank correlations (Wilcox and Muska, 2001) between the different intensities on overlapping temporal windows of 100 ms duration. Because noise in the recordings could have blurred the results, windows with irrelevant signal-to-noise ratio (SNR < 3) were not considered. The SNR was defined for each temporal window as the ratio between the mean of the signal power and its variance. The power rather than the raw signal was used in the definition to avoid negative voltage values. The analysis window covered the 1100 milliseconds period following stimulus onset. The delay for each 100 ms window and for each electrode was selected as the one for which the maximum significant (p<0.01) correlation was obtained. To obtain summary values across subjects we computed the Fisher transform of the correlations and then averaged out over subjects. Note that while the Fisher transformation can fail to provide the correct statistics when the correlations are computed using the Pearson's correlation it is as accurate as bootstrap when more robust non parametric measures of correlations such as the Spearman's rank are used. The significance of the correlation R at the individual means and GM levels was computed in Matlab by transforming the correlation into a t statistic having n-2 degrees of freedom, where n is the number of time points in the analysis window, i.e. $\approx 100\ (from : (100 - time\ lag) \times 1024\ /\ 1000\ )$. The confidence bounds are based on an asymptotic normal distribution of $1/2 \times \ln{(1 + R/1 - R)}$, with an approximate variance equal to $1/(n - 3)$ (Press, 1989). We corrected the p-value with the product of the number of electrodes (64) and the number of time windows (55). This analysis was done using the original Biosemi reference placed over the occipital cortex and repeated using the average reference to make sure that results were reference independent.

## Results

*Dark Noise concerns half of the investigated population*

Figure 4 depicts the individual RTs and FoS curves as a function of the number of photons sent. The abscissa corresponds to the number of emitted photons and the left and right ordinates to the RTs and the proportion of perceived flashes respectively. Subjects were classified into two categories: DN Subjects (DN), i.e. subjects with Dark-Noise or non-zero false alarm rate and NDN subjects i.e. without Dark Noise, depending on whether or not they report false positive trials (Dark Noise). DN subjects might require lower thresholds to reach decision to the expenses of accuracy as expected from a trade-off between accuracy and speed (Gold and Shadlen, 2007). On the other hand we cannot exclude that NDN subjects could indeed have shown false positive trials if more flashes would have been presented. Thus, this classification is to be understood more as a gradual ordering than a dichotomous classification. As shown later we used this classification to confirm the tradeoff between accuracy and speed. Note that in the plots DN subjects are signaled with a red star containing the false negative rate.

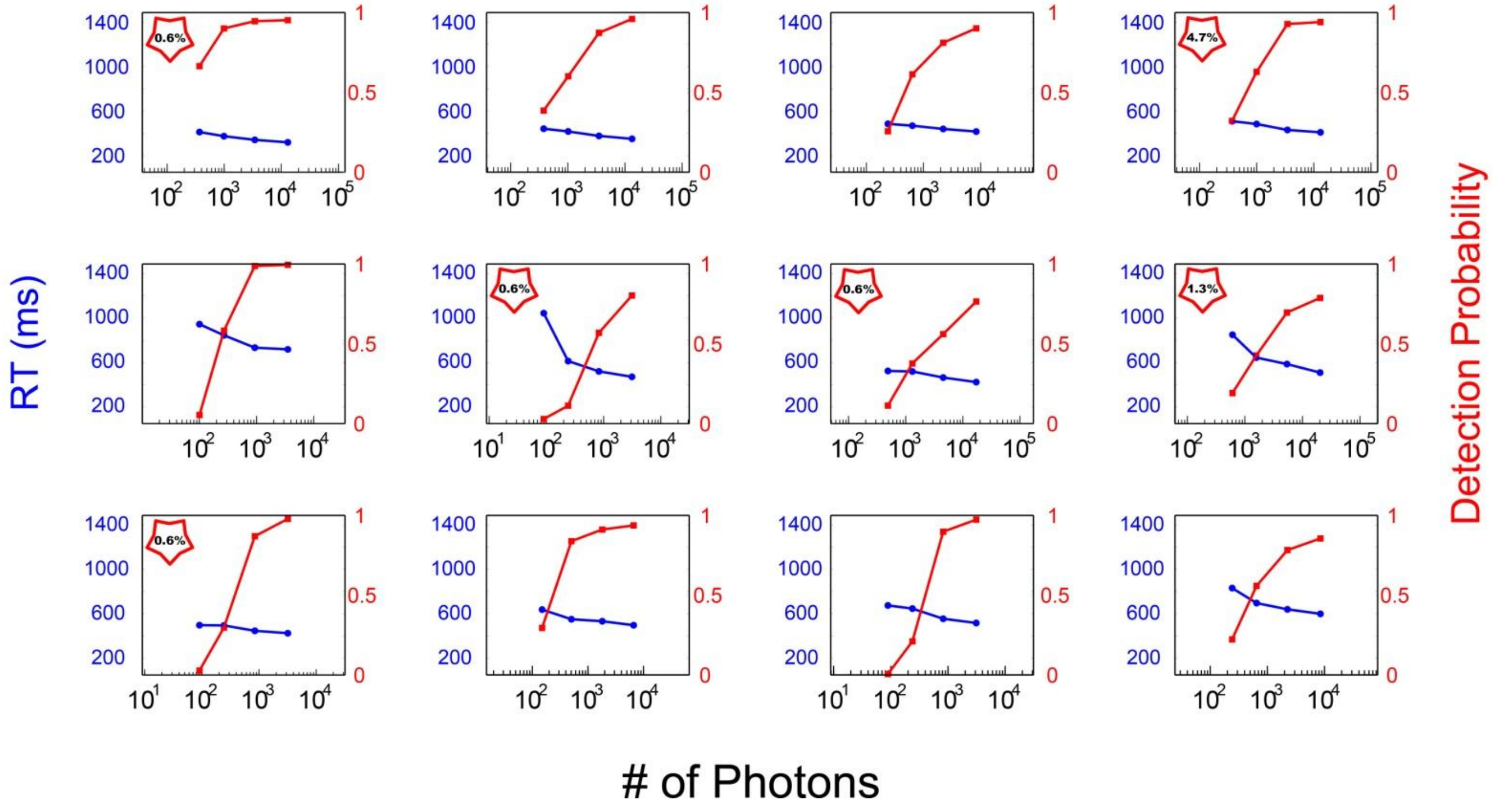

**Figure 4**

As can be seen from Figure 4, exactly half of the participants in the experiment reported seeing flashes when no light was emitted, while the other half showed no evidence of dark noise. Consequently, on the basis of the reduced sample of subjects considered here (N=12), we have to conclude that the probability of observing Dark Noise in a population of young healthy controls is exactly 0.5 and therefore significantly different from zero (binomial test, p==0). This criterion allows for a natural splitting of the subjects according to behavior. However, it is likely that all subjects might exhibit Dark noise if the number of trials increases.

*Frequency of Seeing (Accuracy) and QE*

The estimated number of incident photons necessary to elicit conscious perception determined from the log-Poisson fits to the frequency of seeing curve considerably varied across participants (mean: 815, min 181, max: 3051, std: 903) for a threshold set at 50%. Once these values were corrected by the 0.08 factor estimated by Hecht we obtained: mean: 65, min 17, max: 244, std: 72. The last subject, requiring 244 photons, reported considerable visual fatigue during the experiment. In addition, we observe a non-significant (p=0.11, t-test, p=0.12, ranksum test) but noticeable trend for better performance in young volunteers used to work in the darkness with respect to naïve subjects. This observation, that requires further experimental research, suggests that the efficiency in the conscious detection of dim light might be improved by practice as seen in previous experiment (Hecht, 1942;Sakitt, 1972). This would be hardly the case if visual performances were exclusively dependent on retinal contributions.

*Sensitiviy Index (d')*

Contrarily to the FoS curve which basically reflects the hit rates, the sensitivity index also contemplates the false alarms. Note that the d' index depicted in Figure 5b shows a dependency with the number of incident photons similar to the one in the FoS curve (Figure 5a). This measure allows moreover distinguishing the poorer performances of the noisy subjects.

*DN subjects have smaller sensitivity to specificity ratios (area under the ROC curve) than NDN subjects while sensitivity is similar.*

Both, the mean and the median area under the ROC curve were significantly lower (p = 0.0191, parametric t-test on the mean and p = 0.0022, ranksum non-parametric test on the median) for the group of DN subjects (mean = 0.67 ± 0.05, DN; mean = 0.74 ± 0.01, NDN; median = 0.69, DN; median= 0.75, NDN) than for the NDN subjects. However, no significant differences (p > 0.05 in both cases) between groups were detected on the mean or median true positive rates suggesting that sensitivity is identical in both groups (DN mean = 0.3345, NDN mean = 0.3652, DN median = 0.39, NDN median = 0.4294). Differences in the area under the ROC curve, a composed measure of the trade-off between specificity and sensitivity, indicate that DN subjects are more prone to report false percepts but they are as accurate as non dark noise subjects in detecting similar number of photons. This observation is supported by the statistical comparison of both groups in terms of the number of photons required to elicit conscious perception at 50% of the trials as detected from the FoS curve. Indeed, assuming that only 8% of the emitted photons reach the retina we see that DN subjects require on mean 54 photons to elicit perception in 50% of the trials while noiseless subjects require around 35 photons on mean with no significant differences between groups detected by a parametric (t-test, p = 0.14) or a non-parametric test (rank sum, p = 0.33).

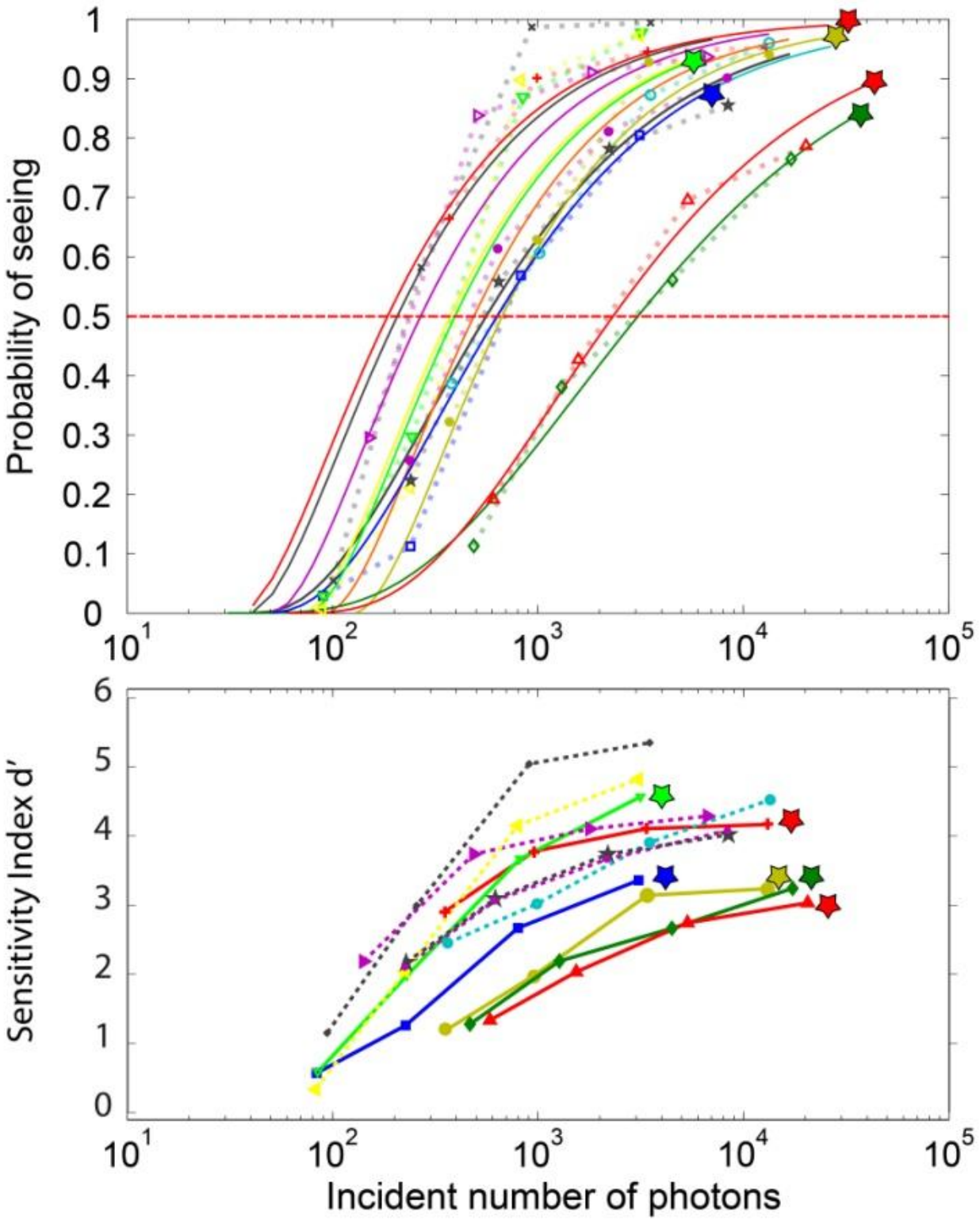

**Figure 5**

*DN subjects are significantly faster to take decisions than noiseless subjects.*

As argued before, in perceptual choice tasks as the one here, experimental evidence indicate that choices are made when the firing rate of selective cortical neural population reach a threshold. Reaction times (RTs) therefore correlate with the time needed to reach the threshold that is dependent on the difficulty of the choice. This effect was clearly observed in our data. The plot of RTs as a function of the intensity of the flashes showed a clear decrease in RTs for increasing intensities at both, the individual (Figure 4) and population level (Figure 6).

As shown in Table 1, the comparison between the distribution of reaction times between subjects with and without dark noise revealed significant differences for all the analyzed intensities. DN subjects were significantly faster to take decisions than NDN subjects. Differences between groups were considerable and varied from 58 ms for the lowest intensity to 100 ms or more for all the others. The rank sum test, as any non-parametric method, is robust to skewed distributions as it makes no assumption about the underlying distribution. In addition, the Lillietest revealed no significant deviations in this dataset from the normality assumption indicating that our RT distributions were not skewed.

**Table 1: Comparison of mean reaction times between DN and NDN subjects. Statistical analyses were performed using parametric (t-test) or non-parametric tests (rank sum). P-values lower than $10^{-6}$ are marked as 0.**

| **Intensity** | **Mean RT DN** | **Mean RT NDN** | **Pval (parametric test, t-test)** | **Pval (non parametric test, rank sum)** |
|---|---|---|---|---|
| I1 | 587 | 645 | 0.03 | 0 |
| I2 | 502 | 621 | 0 | 0 |
| I3 | 474 | 573 | 0 | 0 |
| I4 | 437 | 541 | 0 | 0 |

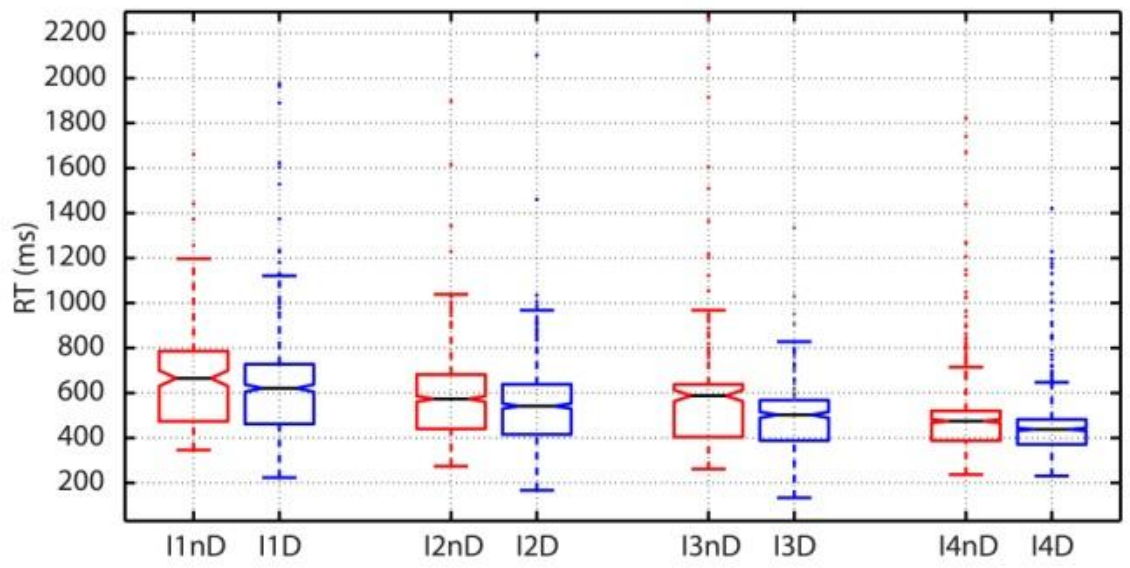

**Figure 6**

*EEG Results*

The grand mean ERPs for the four different intensities are shown in Figure 7. From the figure, it can be readily seen that the early ERP components (N70 and P100) characteristic of visual ERPs in the presence of visible stimuli are absent over contacts placed over primary visual areas, e.g., electrodes Iz and Oz. Moreover, the first significant ($p < 0.01$) deviations in the GM amplitude with respect to a baseline period of 200ms (z-scores >= 2.5 standard deviations away from the baseline mean) were observed first at frontal (Fpz, 211 ms for the strongest intensity), and slightly later for Occipital electrodes (Iz, Oz, 251 ms for the strongest intensity).

The onset, latency and amplitude of the first ERP component recorded over frontal electrodes varied gradually as a function of the intensity of the stimulus. The stronger stimuli (black trace) peaked first than the other intensities. The delay across intensities was more clearly observed on the second, negative ERP component peaking between 400 and 600 ms and appeared over frontal, parietal and occipital electrodes.

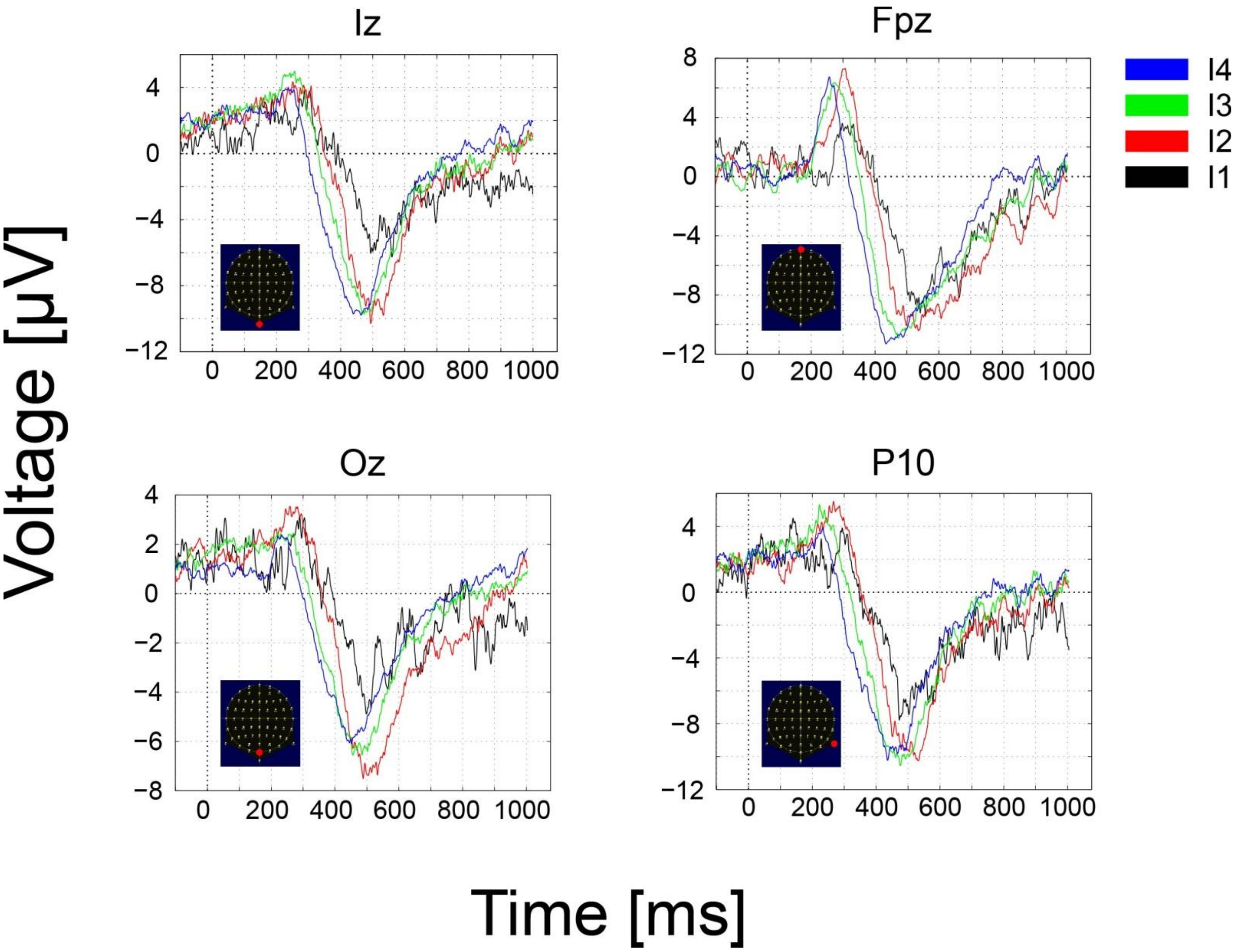

**Figure 7**

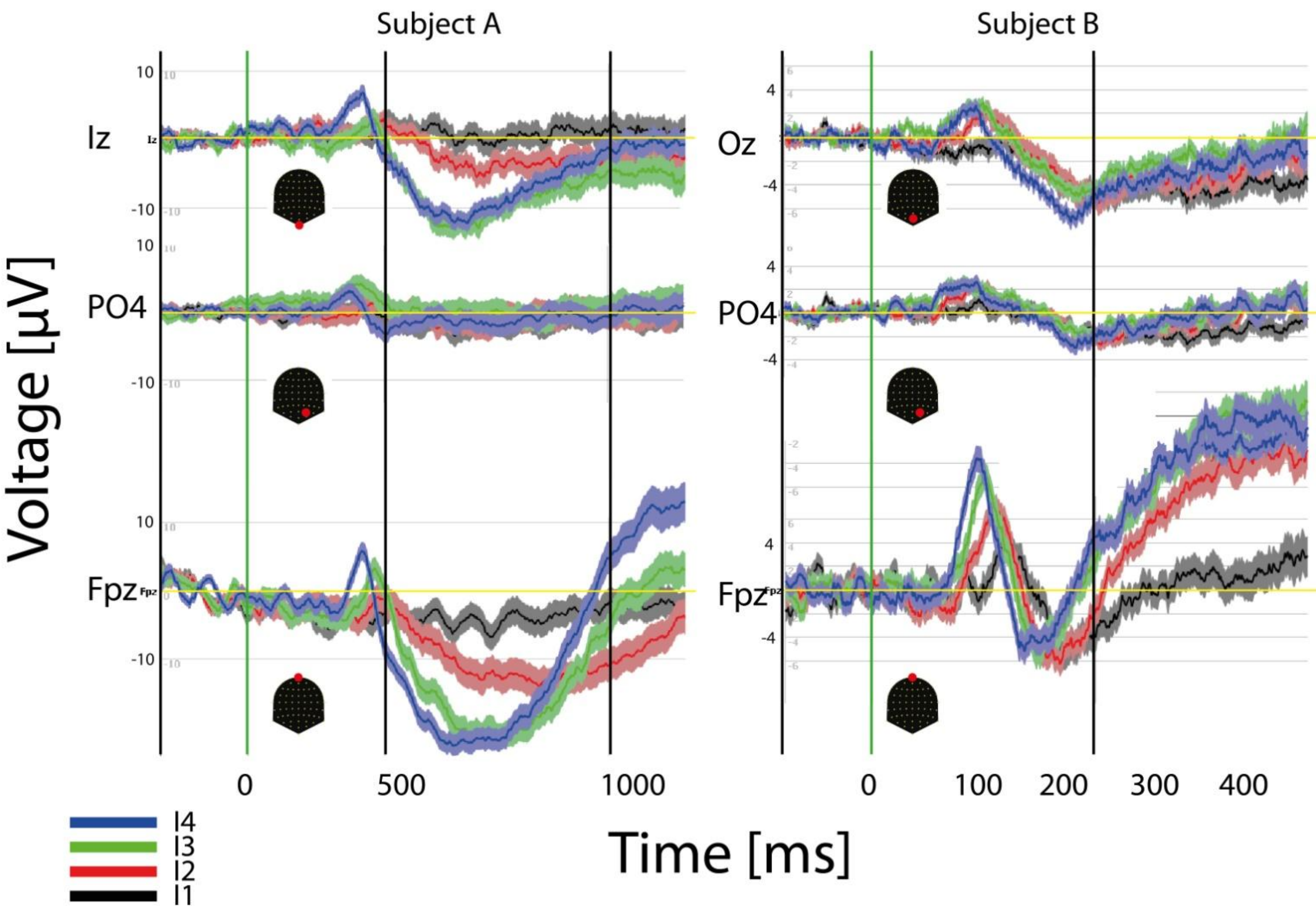

**Figure 8**

ERP results at the single subject level are illustrated in Figure 8. In this case the ERP for the four different intensities (dark thick traces) and the standard error around the mean are depicted for three channels and two different subjects. Each subject is shown in a different column. As observed for the Grand Mean data peak delays across intensities are obvious over frontal contacts (FPz) and much smaller or inexistent over parietal and occipital contacts. Polarities and latencies of the components are similar to those observed for the GM data. Note that occipital responses for the lowest intensity (black trace) are absent at the initial processing stages suggesting that a critical mass of activation of the primary visual cortex necessary to produce the ERP components is absent.

The results of the analysis of the Grand-Mean ERP suggest that the observed behavioral delays in processing the weaker stimuli are mostly introduced by fronto-parietal areas. This observation would be in agreement with the idea that differences in reaction times are due to accumulative effects that involve both retinal and post retinal stages. Our observation is not compatible with decision delays being caused by only longer integration times at the level of the retina. If this were indeed the case the delays should appear at electrodes over visual areas at least as early as in frontal areas. The Grand Mean ERP results therefore suggest that decisional processes play a major role in slowing down responses to weak or faint visual stimuli as suggested by the animal literature. Nevertheless, the analysis of the GM data obscures the contribution of individual subjects and is certainly sensitive to the variability in reaction times observed across subjects. Moreover, this simple ERP analysis is likely to miss latency differences between small ERP components from early visual areas as it has been observed that the intensity of early visual responses in primates V1 decreases with the contrast of the stimuli although the latency of responses increases (Chen et al., 2008).

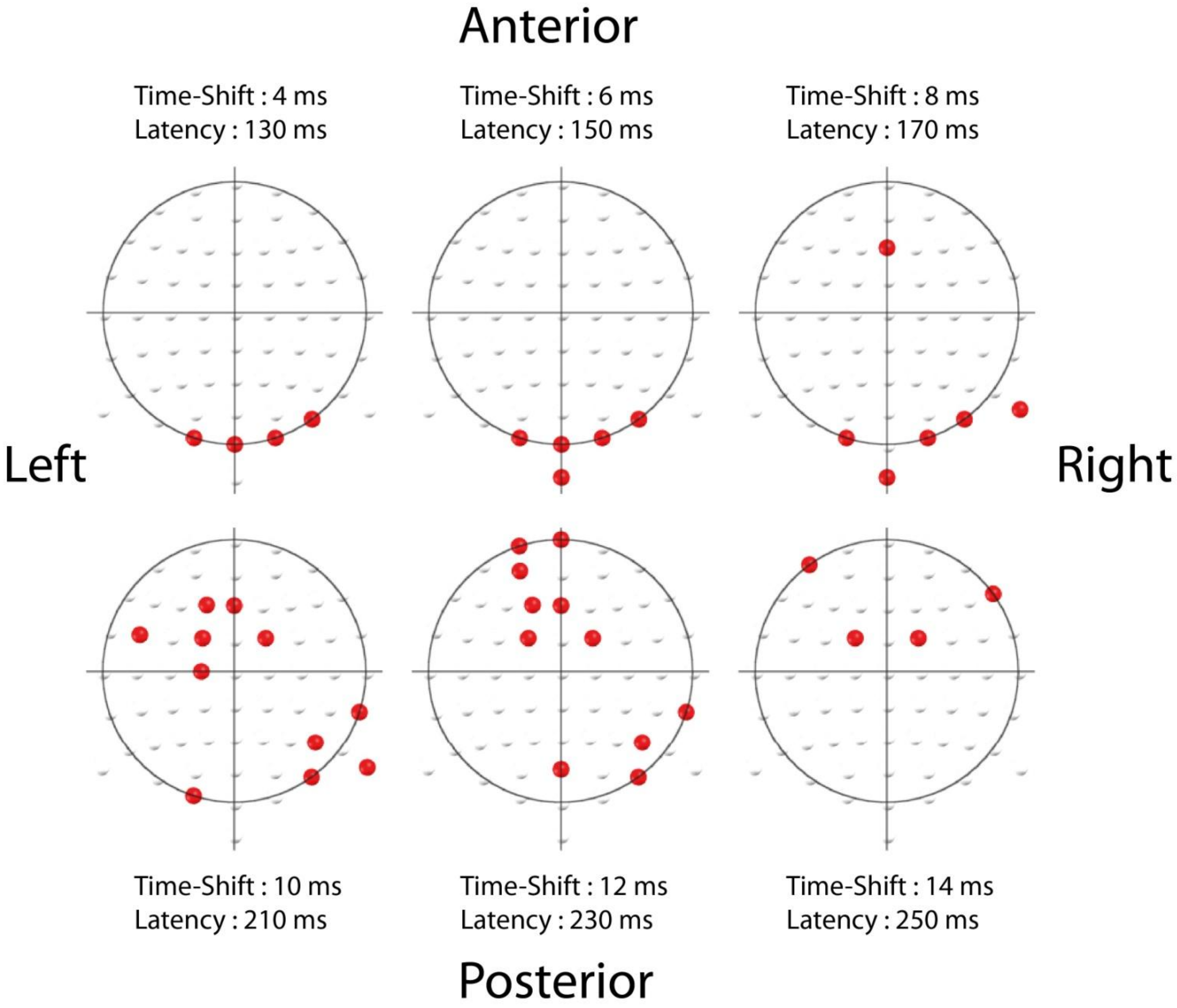

**Figure 9**

To obtain summary statistics over subjects while accounting for interindividual variability we relied on the cross-correlation analysis described at the end of the methods section. In this analysis the goal is to investigate at the individual level, the electrodes showing significant delays in neural responses and to obtain summary statistics once the inter individual differences are accounted for. This method has the advantage of highlighting potentially significant correlations between weak signals that might be overlooked with the more traditional approach based on Grand Mean ERP. The statistical results for this analysis are shown in Figure 9. In the picture, electrodes showing the earliest significant correlation ($r > 0.6$, p-value $< 0.005$) for a given temporal shift (delay) between two intensities (I4 and I3) are indicated by red dots.

As seen from the picture, the electrodes showing the earliest significant correlation between the two higher intensities I3 and I4 vary as a function of the time delay tested. For short temporal delays, signal shifts of approximately 5ms the more significant correlations appear over occipital electrodes. When the time shifts increase to ~10-20 ms, the more significant correlations progressively shift to frontal electrodes. The latency at which electrodes show the earliest significant correlation also covariates with the shift in the signals. For instance, the earliest significant correlation detected across intensities changes from 130 ms for shifts of 4ms to 250 ms for the 14ms delay. Importantly, in the analysis described here we exclusively reported the electrodes for which the earliest significant differences are seen to minimize the potential influence of feedback information (top-down) from high level visual areas into primary visual areas (Bullier, 2001) While not shown here, we observed that delays of approximately 15 ms first seen in parietal areas were later observed in visual areas. The results of the cross-correlation analysis can be then summarized as follow: Small but significant delays of approximately 5 ms across intensities are initially seen over occipital electrodes. Delays become longer (~14ms) and shift to middle central and then to frontal areas when we approach the response times. Consequently, the results are more compatible with a non-linear accumulation of delays across the different processing stages - from perception to decisions - than with a pure retinally induced delay.

While from the scalp topography it is impossible to infer the precise localization of the generators of delays, the very small yet early delays observed over occipital contacts such as Iz, Oz, O1 and O2 (systematically linked to the earliest visual responses) are insufficient to explain the much longer differences in RTs of about 30ms observed for example between intensities I3 and I4. The small early differences in visual areas confirm that a component in RT variability is due to longer integration times at the level of the retina and primary visual areas as already suggested by previous studies ((Pins and Bonnet, 1996;Chen et al., 2008)) and phenomena like the Pulfrich and Hess effects. However, our results strongly suggest that frontal and parietal networks, traditionally linked in invasive studies to decision making processes, are much more involved in the variability of the response times. Noteworthy, while visual responses are hard to evoke and measure using the scalp EEG for such small number of photons, we believe that the results of the analysis described here, and in particular the early detection of the small shifts over occipital electrodes corroborates the sensitivity of our approach to weak signals. Indeed, our results are consistent with animal data which suggest that: 1) neural responses in the lower order neurons are too reliable to account for variability in decision time (Schall, 2002) and 2) neural responses in primates from ganglion cells (Croner et al., 1993 ;Sun et al., 2004) to V1 in alert monkeys (Gur et al., 1997;Gur and Snodderly, 2006) are highly stereotyped in response times. Consequently, animal and human data agree to pinpoint cortically mediated post sensory processing as the major source of variance in RT (Cao and Pokorny, 2010 ).

## Discussion

In this study, we investigated the quantum efficiency (QE) of the human visual system to further clarify: 1) the sources affecting the QE, 2) the divergence between experimental (in animal) and behavioral (in humans) estimates of QE, 3) the origin (retinal or post-retinal) of processes limiting behavioral sensitivity such as the so-called "dark noise", i.e., observers who occasionally report seeing a flash even when none is delivered.

The QE can be estimated from the frequency of seeing curves (FoS) following Hallet (1969). Flashes of light, with a controlled probabilistic distribution of photons were sent into the pupil of 12 dark-adapted subjects who were prompted to indicate if they perceived or not a light. EEG and reaction times were measured along the experiment. The detection threshold, i.e., the number of photons required to trigger a conscious percept, was determined by measuring the fraction of trials in which a flash was reported as perceived as a function of the number of photons incident at the cornea. Despite considerable interindividual variability, we concluded that on mean around 70 photons are required for untrained subjects to trigger perception 50% of the time even though photoreceptors can react to single photons.

Therefore, in agreement with previous experiments the QE estimated from behavior is very low compared with the absorptive QE estimated from the properties of light photoreceptors at the retina. Either much more photons are required to elicit conscious perception than to elicit responses in photoreceptors or dark noise increases the rate of false-positive affecting the sensitivity of the detection threshold and the reliability of the behavioral QE.

The use of behavioural and neurophysiological measures added to the FoS curve allowed to shed further light on the origins of the dark noise. Indeed, exactly half of the subjects showed this effect indicating that behavioral sensitivity considerably varies across individuals. As argued in (Rieke and Baylor, 1998) the dark noise might be the consequence of Poisson fluctuations in photon absorption at the level of the retina. Indeed, continuous noise in mammalian rods can generate fluctuations that look very much like true photon responses (Baylor, 1984). If Poisson fluctuations were generating the dark noise one would expect randomly generated action potentials (APs) at the level of the retina that are transferred to the cortex generating the false percept. Moreover, the generation of each AP is independent of all the other spikes as noise is not expected to simultaneously arise in all retinal cells simultaneously. In this case spikes would be completely described by a random Poisson process. This would imply that all subjects should exhibit the same probability of randomly generated action potentials during the experiment.

However, in a Poisson process the probability of observing random spikes, which could be interpreted as false percepts, increases with the length of the interval, i.e. with the time between the stimuli onset and the decision. It is therefore surprising that DN subjects are significantly faster when in fact; if dark noise were the consequence of Poisson fluctuations at the level of the retina we should expect the converse effect.

Nonlinearities at the retinal level can clearly contribute to the discrepancy between physical QE and perceptual QE. While we cannot fully justify our answer with the available data we believe that retinal nonlinearities alone cannot fully explain the observed differences. Indeed, the trial by trial variability within the same observer for identical number of photons is hard to explain on purely retinal basis. As the physical parameters of the stimuli and that of the cornea are invariant across trials the only possible alternative left is that thermal noise is leading to the trial by trial variability. Yet, the influence of thermal noise seems to be reduced when the number of photons is increased as detection becomes much more accurate.

An alternative is to consider that a significant amount of noise is introduced post-retinal photoreceptor. For instance, dark noise, or at least part of it, could arise during the neural decision making process and as the result of speeded decisions. Indeed, our results fit better to this hypothesis. According to current decision models in cortical areas (Gold and Shadlen, 2007), the accuracy of decisions is inversely proportional to decision time. Yet, given the low false-positive rate of DN subjects (< 5%), it would be unlikely that dark-noise events leading to true positive detection could have significantly decreased the mean reaction time of DN subjects (see Figure 6).

There is growing evidence from primate neurophysiology indicating that in reaction time tasks, a perceptual choice is made when the firing rate of a selective cortical neural population reaches a threshold (Lo and Wang, 2006). Reaction times (RTs) therefore correlate with the time needed to reach the threshold that is dependent on the difficulty of the choice (Smith, 2004). Weak or uncertain stimuli lead to slowly varying accumulation of evidence and longer decision times while certain/strong stimuli lead to quickly growing accumulation of certitude that is reflected in a sharp build up of neural activity that quickly reaches the necessary threshold to reach decisions speeding up the reaction times. It has been also suggested that the threshold can be tuned to optimize the trade-off between speed and accuracy (Smith, 2004).

The fact that there are no differences in sensitivity between both groups of subjects combined with the significantly shorter RTs for the DN individuals is compatible with post-retinal contributions to the decision processes according to current models. This view is supported by the well known perception of phosphenes, i.e. false percepts of light or Dark Noise, after Transcranial Magnetic Stimulation over the occipital cortex. Interestingly, conscious phosphene perception does not appear to be a phenomenon local to the occipital cortex but rather a consequence of extensive recurrent processing (Taylor et al., 2010). Existing experimental evidence therefore suggest that noise leading to false percepts might arise at the level of the visual cortex, extrastriate cortex  or be induced at any later decisional stage that provides feedback to primary visual areas. Consequently, dark noise does not appear to be purely driven by retinal contributions.

The analysis of the EEG recordings additionally suggests that post-retinal contributions play a major role in the decrease in sensitivity measured from the FoS curves. First, the size of the neural signals over primary visual areas was practically zero for the lowest intensity suggesting that a critical mass of neural activation was not reached in the early processing stages. Second, delays in neural responses as a function of the intensity of the stimuli – comparable with behavioral delays – were found at electrodes clustered around frontal and fronto-central regions that have been traditionally linked to decision making. This result coincides with animal electrophysiology suggesting that perceptual choices are driven by the activity of selective cortical population in decision areas.

There are several limiting aspects in this study. First, we are lacking a thorough estimation of the psychometric functions since only four intensities were tested per subject. Yet, increasing the number of intensities per subject would have lead to a hardly tolerable duration for the experiment impacting the reliability of the EEG signals and the behavioral responses. Second, we were not able to achieve a reasonable number of false alarms at the individual level as to build a robust mean ERP for these events. While the number of trials is largely sufficient for robust ERPs for hit trials this is not the case for the false alarms that amount to a few trials in just half of the sampled population. We were therefore unable to investigate in more detail the neural origin of these events. Third, we did not measure RTs for the missed trials. This decision was taken because RTs and false alarms are known to increase with the number of choices available to the participant. However, the nature of false alarms is different in latest case as they are often due to mistakes in button presses. This is a serious concern when working in full darkness. Therefore, in order to identify false alarms strictly linked to false percepts we preferred to keep the choices to a minimum which certainly renders many interesting analysis impossible.

In summary, our results are consistent with a major contribution of post-retinal processing to the QE when measured in terms of the FoS curve. In support of this idea, we observed 1) an inverse dependency between reaction times and the intensity (number of incident photons), 2) Significant differences between mean RTs corresponding to different stimulus intensities for all subjects, 3) Significant differences in RTs between DN and NDN subjects with DN subjects being significantly faster, 4) delays in perceptual decisions are explained by delays in mean ERP responses localized to fronto-central areas rather than primary visual areas. The fact that delays are explained by fronto-parietal areas, known to be involved in decision making reinforce the idea of post-retinal contribution to deficits in sensitivity. Our results are then compatible with Piéron's law that states that, at a constant intensity of the background, reaction times are inversely proportional to the intensity of the target stimulus and independent of the sensorial system (Piéron 1913; Pins and Bonnet 2000). They are also in agreement with the early visual processing described in (Pins and Bonnet, 1996) and results from invasive recordings (Bell et al., 2006) showing that increasing visual stimulus intensity reduces the onset latency of visual responses in several areas outside the primary visual system such as the superior colliculus.

The main conclusion of this study is then that the absorptive QE estimated from the properties of light photoreceptors at the retina is pretty accurate and the human eye is one of the best existing light detectors. However, the QE estimated from behavior is very low as the detector is composed from two different sensors: the retina and the brain. The brain adds a large contribution to the decrease in sensitivity and the losses in QE of the whole system. Low evidence thresholds seem to appear in some subjects that reach faster decisions to the expenses of perceptual errors such as dark noise. In summary, the FoS curves considerably underestimate the QE of the human eye due to post-retinal neural noise that add to fluctuations at the level of the retina.

## References

Aho, A.C., Donner, K., and Reuter, T. (1993). Retinal origins of the temperature effect on absolute visual sensitivity in frogs. *The Journal of Physiology* 463, 501-521.

Alpern, M. (1987). A note on the action spectrum of human rod vision. *Vision Res.* 27, 1471-1414.

Barlow, H.B. (1956). Retinal noise and absolute threshold. *J Opt Soc Am* 46, 634-639.

Baylor, D., Lamb, Td, Yau, Kw (1979). Responses of retinal rods to single photons. *J Physiol.* 288, 613-634.

Baylor, D.A., Nunn, B., Schnapf, J.L. (1984). The photocurrent, noise and spectral sensitivity of rods of the monkey Macaca fascicualris. *J. Physiol.* 357, 575-607.

Bell, A., Meredith, M., Van Opstal, A., and Munoz, D. (2006). Stimulus intensity modifies saccadic reaction time and visual response latency in the superior colliculus. *Experimental Brain Research* 174, 53-59.

Brunner, N., Branciard, C., Gisin, N. (2008). Can one see entanglement ? *Phys. Rev. A* 78.

Bullier, J. (2001). Integrated model of visual processing. *Brain Research Reviews* 36, 96-107.

Cao, D., and Pokorny, J. (2010 ). Rod and cone contrast gains derived from reaction time distribution modeling 10.1167/10.2.11 *Journal of Vision* 10

Chen, Y., Geisler, W.S., and Seidemann, E. (2008). Optimal temporal decoding of neural population responses in a reaction-time visual detection task. *Journal of Neurophysiology* 99, 1366-1379.

Croner, L.J., Purpura, K., and Kaplan, E. (1993 ). Response variability in retinal ganglion cells of primates *Proceedings of the National Academy of Sciences of the United States of America* 90 8128-8130

Fawcett, T. (2006). An introduction to ROC analysis. *Pattern Recognition Letters* 27, 861-874.

Gold, J.I., and Shadlen, M.N. (2007). The neural basis of decision making. *Neuroscience* 30, 535.

Green, D.M., and Swets, J.A. (1966). *Signal detection theory and psychophysics.* Wiley New York.

Gur, M., Beylin, A., and Snodderly, D.M. (1997). Response Variability of Neurons in Primary Visual Cortex (V1) of Alert Monkeys. *J. Neurosci.* 17, 2914-2920.

Gur, M., and Snodderly, D.M. (2006). High Response Reliability of Neurons in Primary Visual Cortex (V1) of Alert, Trained Monkeys 10.1093/cercor/bhj032. *Cerebral Cortex* 16, 888-895.

Hallett, P.E. (1969). Quantum efficiency and false positive rate. *J Physiol.* 202, 421–436.

Hecht, S., S. Schlaer, and H.P. Pirenne. (1942). Energy, quanta and vision. *J. Gen. Physiol.* 25, 819-840.

Jensen, A.R. (2006). *Clocking the mind: Mental chronometry and individual differences.* Amsterdam: Elsevier.

Krantz, D.H. (1969). Threshold theories of signal detection. *Psychological Review* 76, 308.

Lo, C.-C., and Wang, X.-J. (2006). Cortico-basal ganglia circuit mechanism for a decision threshold in reaction time tasks. 9, 956-963.

Merten, K., and Nieder, A. (2012). Active encoding of decisions about stimulus absence in primate prefrontal cortex neurons. *Proceedings of the National Academy of Sciences* 109, 6289-6294.

Pins, D., and Bonnet, C. (1996). On the relation between stimulus intensity and processing time: Piéron's law and choice reaction time. *Attention, Perception, & Psychophysics* 58, 390-400.

Press, W.H., Flannery, B. P., Teukolsvky, S.A., Vetterling, W.T., (1989). *Numerical Recipes in Pascal.* Cambridge: Cambridge University Press.

Ratcliff, R. (2002). A diffusion model account of response time and accuracy in a brightness discrimination task: fitting real data and failing to fit fake but plausible data. *Psychon Bull Rev* 9, 278-291.

Rieke, F., and Baylor, D.A. (1998). Single-photon detection by rod cells of the retina. *Reviews of Modern Physics* 70, 1027 LP - 1036.

Sakitt, B. (1972). Counting every quantum. *The Journal of Physiology* 223, 131-150.

Schall, J.D. (2002). Decision Making: Neural Correlates of Response Time. *Current biology : CB* 12, R800-R801.

Sekatski, P., Brunner, N., Branciard, C., Gisin, N., and Simon, C. (2009). Towards Quantum Experiments with Human Eyes as Detectors Based on Cloning via Stimulated Emission. *Physical Review Letters* 103, 113601.

Smith, P.L., Ratcliff, R. (2004). Psychology and neurobiology of simple decisions. *TRENDS in Neurosciences* 27, 161-168.

Sun, H., Rüttiger, L., and Lee, B.B. (2004). The spatiotemporal precision of ganglion cell signals: a comparison of physiological and psychophysical performance with moving gratings. *Vision Research* 44, 19-33.

Taylor, P.C.J., Walsh, V., and Eimer, M. (2010). The neural signature of phosphene perception. *Human Brain Mapping* 31, 1408-1417.

Thorpe, S., Fize, D., and Marlot, C. (1996). Speed of processing in the human visual system. 381, 520-522.

Wilcox, R.R., and Muska, J. (2001). Inferences about correlations when there is heteroscedasticity. *British Journal of Mathematical and Statistical Psychology* 54, 39-47.

**Figure and table legends**

Figure 1**: Rationale of the experiment and main hypothesis.**

Figure 2**: Experimental Setup.** A light emitting diode (LED) was digitally controlled to produce pulses of light at 500nm wavelength with power varying between 8 pW and 400 pW and very short duration (between 100 and 1ms depending on the subject). The power of the light was individually adjusted to each participant using grey filters and sent trough collimation lens at the end of the fiber to optimally focus the beam to form an angle of approximately 20 degrees with respect to the eye's axis retina where the density of rods is higher. EEG recordings, done using the Biosemi system, were synchronized with the beams onset at the µs level. Around 180 repetitions of four different intensities of light were randomly presented to each participant. In additions an equivalent number of trials (180) were included were participants received a warning signal but no light beam was actually sent into the retina.

Figure 3**: Schematic illustration of the timing of the experiment.** The light pulses were equally randomly distributed within a time windows of 800 ms after the auditory signal. The subjects had 1500 ms to indicate by a button press if he perceived or not the flash.

Figure 4**: Psychometric curves (red) and reaction times (blue) of each subject.** Individual RTs and frequency of seeing curves as a function of the number of photons in the light beam (abscissa). The left and right ordinates depict the RTs and the proportion of perceived flashes respectively. No point is depicted in the graph if the number of flashes reported by the subject as perceived is identically zero when no flash was sent. DN subjects are signaled with a red star, which contains the false-positive rate.

Figure 5**: Top: Psychometric curve for each subjects.** The symbols represent actual data, while the plain lines are their respective log-poisson fits. Each subject is represented in a different color. These fits are used to estimate the minimal number of incident photons required to elicit perception in at least 50% of the trials. The log-poisson distribution allows a reasonable fit for all subjects, even if for a few subjects a linear Poisson distribution may lead to a better fit. The bigger stars at the top of the curves indicate subjects with non-zero false-positive rate. **Lower inset: d-prime sensitivity index for each subject. DN** Subjects with non-zero false-positive rate (plain lines) have a lower d-prime than the NDN subjects.

Figure 6**: Variations in reaction time (RT) as a function of the intensity of the beam** when subjects are divided into two groups according to the presence (DN) or absence (NDN) of false detection (dark noise). The red bars represent the 95% confidence interval. DN = Dark-Noise (Grey), NDN = Non-Dark Noise (White).

Figure 7**: Grand-Mean ERPs over occipital (Iz, Oz), middle frontal (FPz) and parietal contacts (P10) as a function of the intensity of the beam** (the four effective

light intensities). Responses are ordered from the strongest to the weakest intensity (I1 to I4) and the following color convention is used: I4, black; I3, red, I2, green, I1, blue. Note that the onset, latency and amplitude of the first ERP component recoded over frontal electrodes varies as a function of the intensity of the stimulus, i.e., the stronger intensity in black peaks earlier than the other intensities. Delays in neural responses across intensities occur earlier over frontal, and parietal electrodes. The strongest intensities (I4, black) lead to the fastest response and the weakest (I1, blue) to the slowest.

Figure 8: **Individual (2 subjects) ERP averages** for the four intensities at occipital, parietal and frontal electrodes. The highest intensities show faster response at the three sites but effects are more pronounced over frontal contacts.

Figure 9**: Correlation-Based Analysis of intensity-dependant ERP delays :** Schematic top view of the electrodes (black) showing first, among electrodes at a given latency and for given temporal shift, significant correlation ($> 0.6$), p-value ($< 0.005$) and SNR ($> 3$) between intensities 3 and 4.